\documentclass[%
 reprint,
 amsmath,amssymb,
 aps,
 pra,
floatfix,
raggedbottom
]{revtex4-1}
\usepackage{mathtools} 
\usepackage[utf8]{inputenc}
\usepackage{csquotes}
\usepackage[english]{babel}
\usepackage{textcomp}
\usepackage{stfloats}

\usepackage{enumitem,amssymb}
\newlist{todolist}{itemize}{2}
\setlist[todolist]{label=$\square$}

\usepackage{amsmath}
\usepackage{graphicx}

\usepackage{soul} 
\renewcommand{\st}[1]{} 
\renewcommand{\hl}[1]{#1} 

\usepackage[plain]{fancyref}

\usepackage{amstext} 
\usepackage{array}   
\newcolumntype{L}{>{$}l<{$}} 

\usepackage{siunitx}

\usepackage[dvipsnames]{xcolor} 
\usepackage[textsize=scriptsize, backgroundcolor=yellow!20, linecolor=yellow]{todonotes}

\usepackage{float} 

\DeclarePairedDelimiter\abs{\lvert}{\rvert}%

\newcommand{\low}{\textsc{low}}
\newcommand{\high}{\textsc{high}}

\begin{document}
\setstcolor{red}
\title{\bf Optical memories and switching dynamics of counterpropagating light states\\
in microresonators}
\author{Leonardo~Del~Bino$^{1,2,*}$, Niall~Moroney$^{1,3,4}$, and Pascal~Del'Haye$^{1,4}$}

\affiliation{
$^{1}$National Physical Laboratory, Hampton Road, Teddington, TW11 0LW, United Kingdom\\
$^{2}$Heriot-Watt University, Edinburgh, EH14 4AS, Scotland\\
$^{3}$Imperial College London, London, SW7 2AZ, United Kingdom\\
$^{4}$Max Planck Institute for the Science of Light, Erlangen, 91058, Germany\\
$^{*}$Corresponding author: leonardo.del.bino@npl.co.uk
}
\begin{abstract}
The Kerr nonlinearity can be a key enabler for many digital photonic circuits as it allows access to bistable states needed for all-optical memories and switches.
A common technique is to use the Kerr shift to control the resonance frequency of a resonator and use it as a bistable, optically-tunable filter.
However, this approach works only in a narrow power and frequency range or requires the use of an auxiliary laser.
An alternative approach is to use the asymmetric bistability between counterpropagating light states resulting from the interplay between self- and cross-phase modulation, which allows light to enter a ring resonator in just one direction.
Logical \high\ and \low\ states can be represented and stored as the direction of circulation of light, and controlled by modulating the input power.
Here we study the switching speed, operating laser frequency and power range, and contrast ratio of such a device.
We reach a bitrate of \SI{2}{Mbps} in our proof-of-principle device over an optical frequency range of 1 GHz and an operating power range covering more than one order of magnitude. We also calculate that integrated photonic circuits could exhibit bitrates of the order of Gbps, paving the way for the realization of robust and simple all-optical memories, switches, routers and logic gates that can operate at a single laser frequency with no additional electrical power.
\end{abstract}
\maketitle

\section{Introduction}

At the present time, fiber optics telecommunication nodes convert light into electrical signals with fast photodiodes to perform processing, routing, or storing of information.
This is an established technology; however, the increasing traffic on networks is approaching the limitation imposed by the double conversion of the signals from optical to electronic and back to optical again. This conversion requires an additional layer of complex, expensive and power-hungry devices such as lasers and electro-optic modulators (EOMs).
To overcome these limitations, photonic circuits are being studied as a viable alternative to conventional electronic circuits \cite{Shacham2007,Gunn2006}. 

There are several technologies that allow direct control of light. Some are still controlled by non optical inputs such as EOMs, micro-electro-mechanical systems (MEMS) \cite{Abdulla2011} or thermally tuned devices \cite{Chen2018}.
All-optical devices have been theorised \cite{Xu2011, Poon2009, Chai2016} and demonstrated to be a viable solution for all-optical switches using different forms of nonlinear responses \cite{Rios2015, Friberg1987, Takahashi1996, Min2008, Akin2014, Chen2017, Fang2014, Ali2016}
All of these devices rely on a nonlinear optical response that may require significant optical input power to work. When low operating power is required, resonant optical cavities can be used.

Microresonators are a widely used platform for the observation of nonlinear optical phenomena, because of their high Q-factors and small mode volumes. Not only do they enable very high circulating power intensities to be reached, but also their small size makes them ideal candidates for integrating nonlinear effects on chip-scale devices and optical circuits.
Ring lasers \cite{Chen2011,Osborne2009,Hill2004,Liu2003,Liu2010} are one example of devices that enable an effective directional switching, with the lasing direction being controlled by an input seed with response times below the nanosecond level.
However, ring lasers are not passive and require either optical pumping at a different wavelength from the signal or electrical pumping to create population inversion in the active medium.
An alternative approach is to exploit the power dependence of the resonance frequency in nonlinear resonators to realize a fast-tunable switch.
Multiple phenomena change the effective length of the resonator, and therefore its resonance frequency: Kerr effect \cite{Yoshiki2014}, thermo-refractivity \cite{Almeida2004}, thermal expansion, and two-photon absorption (TPA).
However the timescales of the resonator responses are different: the Kerr effect can be considered instantaneous; carrier lifetimes broadly depend on the semiconductor and its doping, and generally can span from sub-picosecond \cite{Currie2016} to almost millisecond; the resonator heating that affects both the refractive index and the cavity length shows multiple characteristic timescales, ranging from the microseconds level to several seconds.
Considering this broad span in response times, and the fact that different nonlinear effects often give opposite contributions to the refractive index change, most solutions aim to maintain their contributions at different magnitudes.
Photonic crystal cavities \cite{Shinya2008, Nozaki2015, Soljacic2002, Tanabe2005, Notomi2005, Nakamura2004} are usually very small, on the order of a few tenths of micrometers, and can switch at \si{GHz} rates.
Semiconductor resonators also represent a very promising approach in this direction \cite{Almeida2004, Wen2012}, being easy to integrate in silicon micro-fabrication techniques. However, in both cases, the interplay between TPA, the Kerr effect and temperature drift hinders their long term stability. 

\begin{figure}[ht]
    \centering
    \includegraphics[width=\columnwidth]{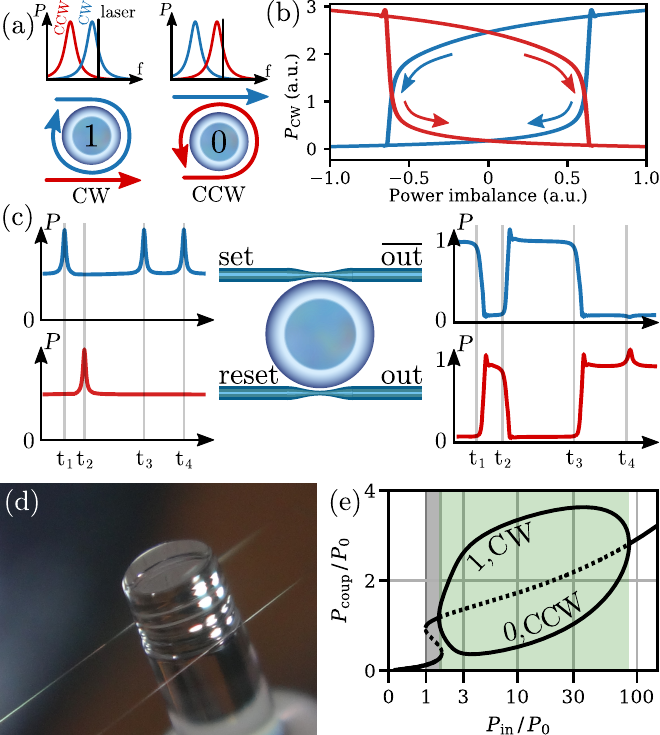}
    \caption{(a) Schematic of clockwise and counterclockwise light states in a microresonator and corresponding mode spectra. \st{Light paths are represented by the arrows. Different colors are used for clarity, despite both directions having the same frequency.}
    (b) Calculation of the circulating power in both directions as a function of the imbalance in the input powers. Note the hysteresis separating the two directional states.
    (c) Timing diagram showing how the device works as a set-reset flip-flop. \st{The input ports `set' and `reset' are modulated around a constant bias power. Pulses above the bias in the input ports determine the subsequent outputs.} Note that at $t_4$ an additional `set' input does not change the output.
    (d) An image of the whispering gallery mode resonator and the two coupling tapered fibers used in the experiment.
    (e) Representation of the typical coupled power profile for the Kerr nonlinearity in case of equally intense counterpropagating beams for a laser blue-detuned by $\delta = 2.5 \,\gamma$\st{, where $\gamma$ is the effective (or coupled) half width half maximum (HWHM) of the resonance. Circulating power is plotted in units of normalisation power defined in fref{eq:dimensionless}.} The dashed curve shows unstable solutions. The two areas with different kinds of bistability are indicated. The green area arises from the interplay between counterpropagating light. Here the resonance frequencies and coupled powers are different for the two directions. The grey area is a symmetric bistability and coupled powers and resonance frequencies are identical in the two directions.
}
    \label{fig:s-shape}
    \vspace{-3mm}
\end{figure}

In this work, we demonstrate a device that exploits the Kerr effect differently.
It has recently been shown that the Kerr effect prevents light of the same frequency, and above a threshold power $P_\mathrm{th}$, from circulating simultaneously in both directions in whispering gallery mode microresonators \cite{DelBino2018, DelBino2017, Cao2017, Copie2019}.
The direction of circulation can be controlled by varying the inputs to the resonator.
In this context, the bistability arises from the twofold contribution to the refractive index change $\Delta n$ of the cross-phase modulation (XPM) compared to the self-phase modulation (SPM).
Therefore, the refractive index change can be different for the two propagation directions, clockwise (CW) and counter-clockwise (CCW):
\begin{equation}
\Delta n_{1,2} = n_2 (I_{1,2} + \textbf{2}\, I_{2,1})
\label{eq:deltaKerr}
\end{equation}
where the subscripts $1,2$ identify the two directions CW and CCW, $I_{1,2}$ are the circulating light intensities in the resonator, and $n_2$ is the nonlinear refractive index.
In an optical cavity with length $L$, the resonance frequency is related to the refractive index $n$ by $ m \lambda = n L$, $m \in \mathbb{N}$. If the resonator exhibits two different effective refractive indices in the two directions, the resonance frequencies will also differ, resulting in in-coupling of light in only one direction. The direction of light propagation in the resonator can be interpreted as digital states \high\ and \low\ or 1 and 0 as shown in \Fref{fig:s-shape}(a).
Furthermore, the hysteresis of this system (\Fref{fig:s-shape}(b)) allows it to act as a set-reset flip-flop, the fundamental component of digital memory.
In other words, each directional state is stable, since the switching between the two states requires overcoming the hysteresis. This is illustrated in \Fref{fig:s-shape}(c) and demonstrated in the results and discussion section.
When tuning the laser into resonance or increasing the laser power, the direction with the higher input power will couple more light into the resonator, shifting the counter-propagating resonance frequency away due to the XPM.
Consequently, the weaker direction is prevented from coupling into the resonator.
Even if the input power is equal in both directions the resonator still undergoes symmetry breaking, resulting in just one of the two directions coupling into and circulating in the resonator.
Thus no indeterminate states can arise when both the laser detuning and the input powers are in the symmetry-broken region \cite{DelBino2017,Woodley2018}.
Since the two states are separated by an unstable region, it is not possible to transition continuously from one state to the other. Instead, the system tends to stay in its current state, unless significantly more power is launched in the counterpropagating direction (overcoming a hysteresis).

In most works that use the unidirectional Kerr effect or two-photon absorption (e.g. \cite{Almeida2004, Shinya2008, Notomi2005}), the laser is red-detuned from the resonance in one of the two bistable states.
The optical and thermal nonlinear effects shift the resonant frequency towards the laser frequency, causing a bistability referred to as the S-curve that exists in a limited region of detuning and launched optical power, shown in grey in \Fref{fig:s-shape}(e).
Instead, in our method, an additional bistability arises from the difference between the SPM and the XPM \cite{Agrawal1987, DelBino2017} for counter-propagating directions.
This counter-propagating bistability exists in a much broader range of input powers and detunings compared to the S-curve (see \Fref{fig:s-shape}(e)).
This allows us to operate the laser blue-detuned from the resonance, exploiting the thermal and Kerr effects, which self-locks the resonance to the input laser. If the input power increases or the laser drifts closer to the resonance frequency, the circulating power is subject to a negative feedback effect that causes the resonance frequency to move away from the laser, stabilizing the intracavity power and the detuning \cite{Carmon2004}. Since all the thermal effects act symmetrically with respect to the direction of light and there is thermal locking we do not include the temperature dependence in our theoretical model.

\begin{figure}[ht]
    \centering
    \includegraphics[width=\columnwidth]{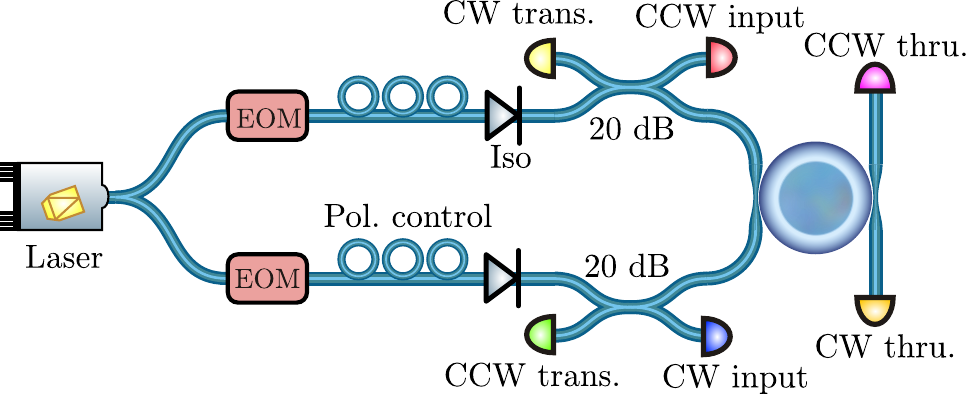}
    \caption{A schematic representation of the experimental setup. An external cavity diode laser (ECDL) is amplified and the output is split equally into two branches. In each branch, the power is modulated by an EOM and the polarization is adjusted to match the resonator mode before coupling the light in opposite directions into the resonator via a tapered fiber. Two isolators prevent the light from traveling back into the laser. A second tapered fiber is weakly coupled to the resonator to monitor the circulating light. Six photodiodes detect the input powers, the transmitted powers through the input fiber, and the circulating powers via the second fiber.}
    \label{fig:setup}
\end{figure}

\section{Methods and Theory} 
The experimental setup is shown in \Fref{fig:setup}. A \SI{1550}{nm} external cavity diode laser (ECDL) is amplified with an erbium-doped fiber amplifier (EDFA) and split equally into two branches that couple light in opposite directions into a microresonator.
The microresonator is made of fused silica (SiO$_2$) with a diameter of 2.7~mm and a Q-factor of $4 \, \times \, 10^8$. \st{Each branch includes a polarization controller to maximize the transmission of the subsequent Mach-Zender electro-optic amplitude modulators (EOMs), which are used to modulate the light power.
Another} \hl{Mach-Zender electro-optic amplitude modulators (EOMs) are used to modulate the light power in each direction.} Polarization controllers allows the alignment of the polarization of the laser to one of the resonator modes, maximizing the coupling efficiency.
An isolator prevents the light from the opposite branch from reaching the EOM and the laser.
A \SI{20}{dB} coupler allows us to tap a fraction of the light to measure the input power to the resonator and the light transmitted from the other branch.
Each branch is connected to one end of a tapered optical fiber with diameter below \SI{1}{\micro m}, which injects light in the microresonator through evanescent coupling.
A second tapered optical fiber is coupled to the resonator to measure the circulating power (see Figure \ref{fig:s-shape}(d)).

The Kerr effect depends on the circulating powers in the CW and CCW directions. These powers can be directly controlled via the input powers into the resonator or via the laser detuning from the resonance.


The time derivative of the field inside the resonator in each direction is:
\begin{equation}
\dot{e}_{1,2} = \tilde{e}_{1,2} - \left[1 + i\left(\abs{e_{1,2}}^2 + 2\abs{e_{2,1}}^2 -\Delta \right) \right]e_{1,2}
\label{eq:time}
\end{equation}
where $\tilde{e}$ is the input field scaled by the coupling efficiency. The losses are proportional to the circulating field $e$ with an imaginary part denoting the detuning of the laser from the ``cold'' resonance and the additional detuning caused by the Kerr effect comprising SPM and XPM.
All the quantities are normalized to be dimensionless: the detuning $\Delta$ is expressed in units of the half linewidth of the resonance $\gamma$ and the power is in units of the normalization power $P_0$.
\begin{equation}
\Delta = {\delta}/{\gamma}; \quad P_0 = \frac{\pi n_0^2 V}{n_2 \lambda Q Q_0}
\label{eq:dimensionless}
\end{equation}
where $n_0$ is the refractive index of the material, $V$ is the effective mode volume, $n_2$ is the nonlinear refractive index, $\lambda$ is the vacuum wavelength of the laser, and $Q$ and $Q_0$ are the coupled and intrinsic Q-factors of the resonator.
The dimensionless input powers into the resonator in directions $i=1,2$ are
\begin{equation}
\tilde{p}_i = \left\vert \tilde{e}_i \right\vert ^2= \frac{\eta P_{\mathrm{in},i}}{P_0}
\end{equation}
where $\eta$ is the coupling efficiency and $P_{\mathrm{in},i}$ is the input power. The dimensionless power coupled into the resonator is
\begin{equation}
p_i = \left\vert e_i \right\vert ^2 = \frac{P_{\mathrm{coup},i}}{P_0} = \frac{2\pi}{\mathcal{F}_0}\frac{P_{\mathrm{cir},i}}{P_0}
\end{equation}
where $\mathcal{F}_0$ is the intrinsic finesse of the resonator and $P_\mathrm{cir}$ is the optical power circulating in the resonator. We introduce the quantity $P_\mathrm{coup}$ because it has the same order of magnitude of the input power and, in a steady state, it can be measured as the difference between the input and the transmission in the input tapered fiber.
The stationary state of \fref{eq:time} corresponds to the coupled Lorentzians model analysed in \cite{DelBino2017,Woodley2018}:
\begin{equation}
p_{1,2}=\frac{\tilde{p}_{1,2}}{1+\left(p_{1,2} + 2 \, p_{2,1} - \Delta \right)^2}
\end{equation}
In case of symmetric pumping, the threshold power for the symmetry breaking \cite{DelBino2017}, i.e.~the beginning of the green shaded area in \Fref{fig:s-shape}(e) is given by:
\begin{equation}
P_\mathrm{th} = \frac{8}{3\sqrt{3}} P_0 \simeq 1.54 \, P_0
\label{eq:Pth}
\end{equation}
Our model assumes the intensity profile in the cavity to be spatially constant across the ring since only one cavity mode in each direction is being excited.
To ensure that this is the case in the experiment, we \st{optimise the geometry of the resonator such that the dispersion is} \hl{choose a spatial mode with geometric dispersion} unfavorable to any kind of modulation instability \hl{at low power} or other nonlinear effects that generate light in different modes.

The power is modulated anti-symmetrically in the two branches to keep the total input power in the resonator constant; i.e.~if the input power in one direction is increased, the power in the other direction is reduced by the same amount.

\onecolumngrid

\begin{figure}[H]
    \centering
    \includegraphics[width=1\textwidth]{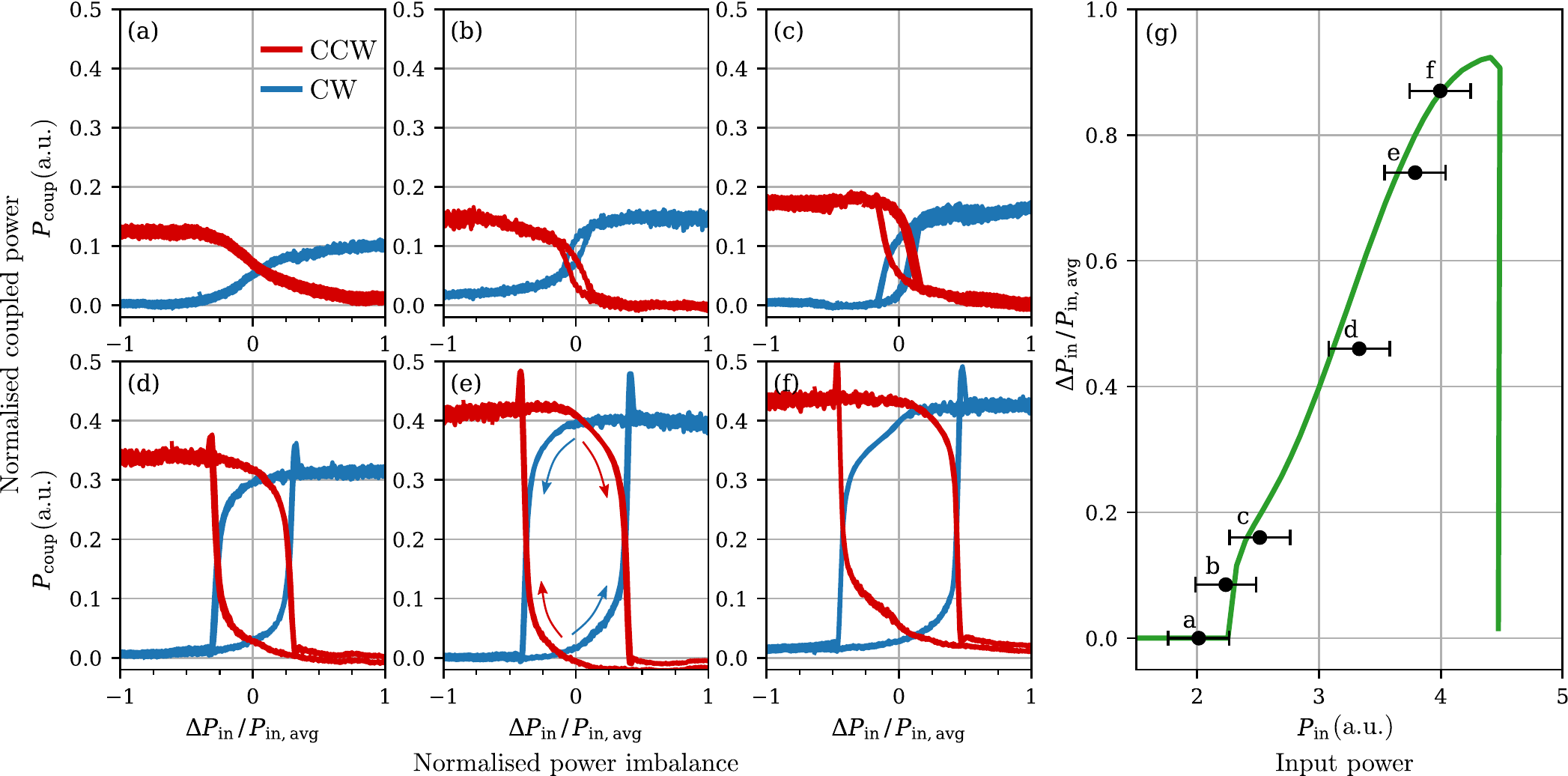}
    
    \caption{Measurement of the hysteresis amplitude for different circulating powers. \hl{The circulating power is changed via the laser detuning.} In panels (a-f) on the left, the input powers in the two directions are anti-symmetrically modulated with a triangular waveform at $\sim4$~kHz and the corresponding coupled power in both directions is plotted as a function of the power imbalance. The black data points in panel (g) show the hysteresis amplitude measured as the relative power imbalance needed to cause switching for the different coupled powers of panels (a-f). The green trace shows the corresponding theoretical model. See \hl{Figure S2} in the Supplemental Material for more details.}
    \label{fig:hysteresis}
\end{figure}
\twocolumngrid
Therefore, the effective detunings
\begin{equation}
\Delta_\mathrm{eff \, 1,2} = \abs{e_{1,2}}^2 + 2\abs{e_{2,1}}^2 -\Delta_\mathrm{laser} 
\label{eq:delta_eff}
\end{equation}
are just switched between two steady-state values, with one taking one value when the other takes the other. In mathematical terms this corresponds to switching the indices $1 \leftrightarrow 2$ or $\mathrm{CW} \leftrightarrow \mathrm{CCW}$.
This method also avoids thermally-induced resonance shifts that may distort the switching profile \cite{Carmon2004}.
However, the hysteresis between the two states is still present when modulating just one direction as long as the total power remains in the symmetry-broken region.
\section{Results and Discussion}


We define the hysteresis amplitude as the power imbalance that is needed to induce state switching in either direction.
To measure this we anti-symmetrically change the powers in the two directions, keeping the total power constant. The power ramp has a period much slower than the switching time. In the meantime, we monitor the input and transmitted powers in the two directions as shown in \Fref{fig:hysteresis}.
This measurement is repeated for different in-coupled powers, which is achieved by changing the laser detuning in this measurement.
  
To better visualize the hysteresis amplitude, Figures~\ref{fig:hysteresis}(a-f) show the coupled power in each direction versus the input power imbalance (normalized by the average input power) as the laser is tuned down in frequency towards the resonance. The measurements shown in \Fref{fig:hysteresis} span a range of detuning of about \SI{10}{GHz}. For larger detunings, the thermal effect cannot follow the laser frequency and the resonance jumps back to its ``cold'' state, away from the laser. 
When the laser is far blue detuned from the resonance and the\st{z} coupled power is small there is no hysteresis and the coupled power change continuously between the two counter-propagating states. As $P_{\mathrm{coup}}$ increases, the hysteresis grows wider and the switching appears sharper.
The overshoot observed in panels (d-f) is a dynamical effect that occurs over a timescale of the order of the cavity lifetime and is explored in detail later on in this paper. 

This behavior is well predicted by the theory, as shown in panel (g). The green curve shows the hysteresis amplitude in terms of relative power imbalance as a function of the input power, calculated using a time step simulation of \Fref{eq:time} with the same parameters as used in the experiment. The black dots represent the measurements of hysteresis amplitude for panels (a-f).
\begin{figure}[h]
    \centering
    \includegraphics[width=0.85\columnwidth]{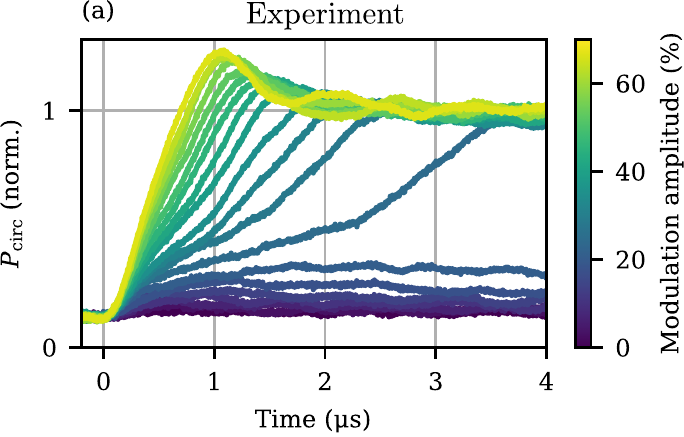}
    
    \vspace{1 mm}
    
    \includegraphics[width=0.85\columnwidth]{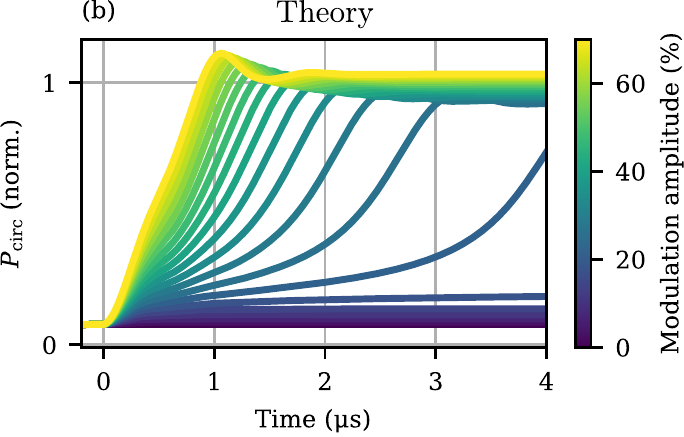}
    \caption{Switching profile at the output port for different modulation amplitudes: (a) data and (b) numerical simulation. The resonator is set in the \low\ state, then the input powers are changed at $t=0$ by an increasing amount indicated as percentage of the average power. The circulating power is normalised to the steady state value.}
    \label{fig:profile_vs_modamp}
\end{figure}
Note that the hysteretic behavior is observed over a broad range of optical input powers.
The lower end of this range is the threshold power $P_{\mathrm{th}}$.
As the input power is increased, the hysteresis increases in width up to about $5P_{\mathrm{th}}$ when four-wave mixing effects, such as modulation instability, start to appear \hl{despite the adverse dispersion conditions}.

The operational \hl{optical} frequency range \hl{around a resonance} is dependent on the input power, spanning from \SI{1}{GHz} for input powers just above $P_{\mathrm{th}}$ to reach \SI{10}{GHz} at higher input powers, allowing free-running operation for several hours. This can be extended by stabilizing the coupled powers via the transmission signal.

We now analyze the switching profile, i.e.~how the monitor tapered fiber signal changes over time during the switching. In \Fref{fig:profile_vs_modamp}(a), the resonator is set in the \low\ state with equal input powers in both directions. At $t=0$ the input power is increased in the measured direction by a percentage indicated in the color-bar and decreased by the same amount in the other direction.

A corresponding theoretical calculation is shown in \Fref{fig:profile_vs_modamp}(b). The simulation starts from the \low\ symmetry-broken steady state and the input field amplitude $\tilde{e}_1$ and $\tilde{e}_2$ in \Fref{eq:time} are set to a common value corresponding to the experiment at $t=0$. The evolution is calculated using a fourth-order Runge-Kutta method with time steps of $1/(100 \, \gamma)$.

For modulations under $25\, \%$ of the average power, the signal barely changes because the modulation is not sufficiently large to overcome the hysteresis.
For modulation amplitudes above $25\, \%$ the switching speed gets increasingly faster until it saturates at about \SI{500}{ns}, which corresponds approximately to the cavity rise time (see Equation \ref{eq:lifetime}).

The switching speed is determined both by the physical characteristics of the device, such as Q-factor, material and mode volume, as well as the input.
The average input power and the detuning between the laser and the resonator play only a minor role in the switching dynamics.
Indeed, once the power is above the threshold power $P_{\mathrm{th}}$ and the detuning is in the range of bistable behavior, the switching profile is barely affected by these parameters (see Supplemental Material for further analysis).
The speed of the dynamics is intrinsically limited by the cavity \hl{lifetime}, i.e.~the higher the Q-factor of the resonator, the longer it takes the light to build-up in the cavity or to leak out of it. This speed is expressed with the following quantities, each one popular in a specific field.
\begin{equation}
\tau_c = \frac{Q}{2 \pi \nu_0}; \qquad \tau_\mathrm{rise} = 2.197\, \tau_c; \qquad R=\frac{1}{\tau_\mathrm{rise}}
\label{eq:lifetime}
\end{equation}
where $\tau_c$ is the cavity lifetime\hl{, i.e. the time it takes for the cavity to leak $1/e$ of the circulating power, an alternative definition is the $10\,\% - 90\,\%$ power rise time $\tau_\mathrm{rise}$ that is mathematically related to $\tau_c$. We define the bitrate $R$ as the inverse of the rise time.}
The silica rod resonator used in this work has a coupled Q-factor of \num{2e8} (see \hl{Table I in the Supplemental Material}) and is expected to have a rise time of \SI{360}{ns} limiting the bitrate ($R$) to about \SI{3}{Mbps}.
In our case, the switching process is slightly slower than the cavity \hl{lifetime} due to the interplay between the Kerr-related detuning and the intracavity power.
The choice of a high-Q-factor resonator allows us to consider the EOM modulation as instantaneous.
Hence, the input can be approximated as a Heaviside step function and the response measured directly, without having to perform any deconvolution.
\begin{figure}[H]
    \centering
    \includegraphics[width=0.9\columnwidth]{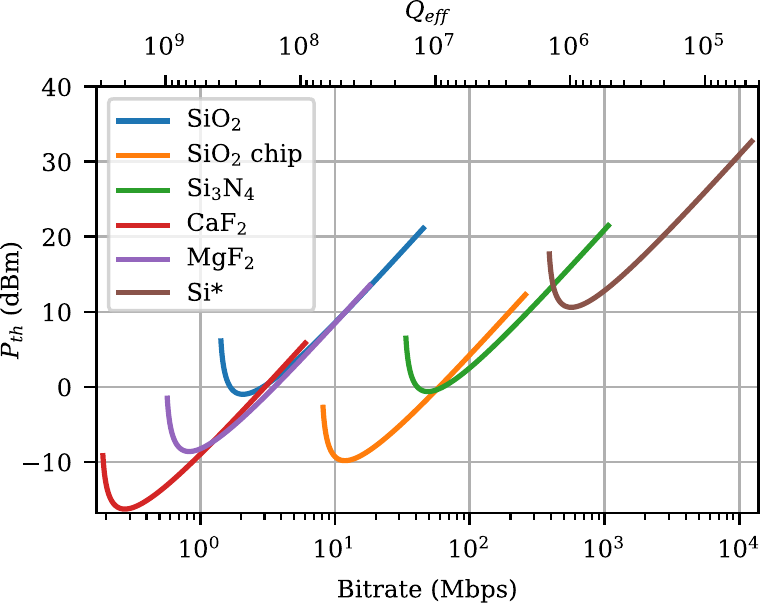}
    \caption{Power requirements and corresponding bitrates for different materials. Each line represents the possible trade-off that can be obtained by changing the coupling between waveguide and resonator. The parameters used in this plot and their relative sources are reported in the SM. * $Q_\mathrm{eff}$ scale is not valid for Si since it works at a different wavelength.}
    \label{fig:Nike}
\end{figure}
\onecolumngrid

\begin{figure}[H]
\centering
  \includegraphics[width=0.245\columnwidth]{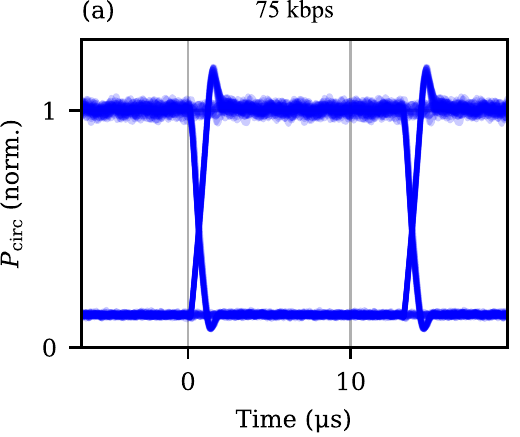}
  \includegraphics[width=0.245\columnwidth]{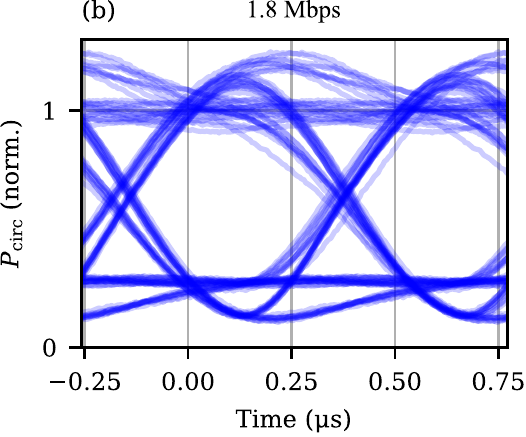}
  \includegraphics[width=0.245\columnwidth]{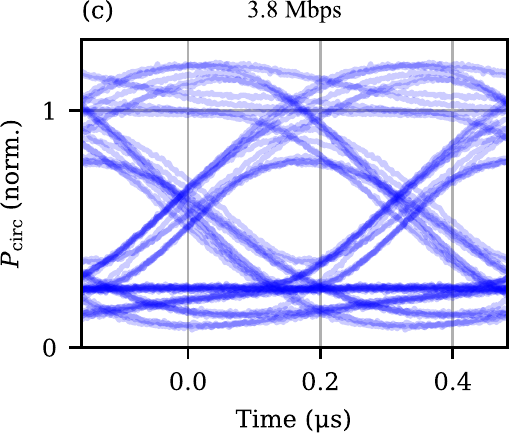}
  \includegraphics[width=0.245\columnwidth]{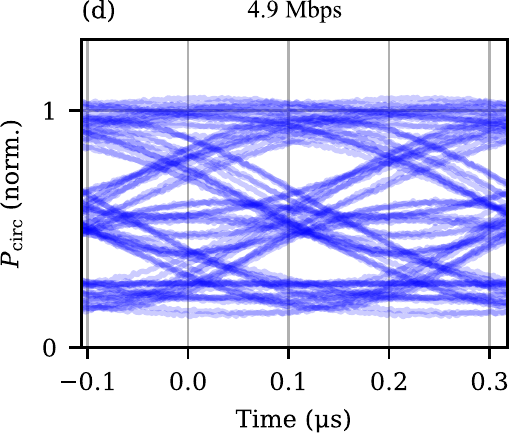}
  \vspace{-5mm}
\caption{Eye diagrams. Response of the resonator to a random bit pattern with anti-symmetric input power modulation. The bitrates in the four panels are: (a) 75~kbps, (b) 1.8~Mbps, (c) 3.8~Mbps, and (d) 4.9~Mbps. The acquisition is synchronized with the clock transition. The power, normalized to the steady-state \high\ value, is measured at the output fiber. Only the CCW direction is shown for clarity. \hl{See Supplemental material for more details.}}
    \label{fig:eye}
\end{figure}
\twocolumngrid
In addition, it is important to highlight that the threshold power for bistability scales as $1/Q^2$ as shown in Equations \ref{eq:dimensionless} and \ref{eq:Pth}.
By achieving $P_\mathrm{th}$ on the order of 1~mW, high-Q-factor resonators thus allow us to explore the regime $\tilde{p}_i \gg 1$, where the input power is much higher than $P_\mathrm{th}$, without being limited by the EOMs maximum operating power.

If higher switching speed is required, the Q-factor must be lower. One way of achieving this is to tune the coupling of the monitor tapered fiber to introduce more losses, and hence a lower $Q$. However, this would increase the power required to achieve the bistable regime.
Thus, to achieve faster operation at low power the other parameters in \Fref{eq:Pth} need to change.
In particular, smaller resonators made of highly nonlinear materials such as silicon nitride reduce the power required by the device. 
A typical silicon nitride waveguide ring resonator would result in a device operating with the same power requirements but with speed over \SI{1}{Gbps}. Additional simulations of how the bitrate and power requirements would vary on different platforms is shown in \Fref{fig:Nike}.
Note that each material has a $P_\mathrm{th}$ corresponding to the coupling that provides optimal power efficiency. For lower coupling, less light enters the resonator, meaning that higher input power is required. On the other hand, higher coupling degrades the coupled Q-factor, allowing faster switching at the expense of higher power requirements.

\Fref{fig:eye} shows an eye diagram, providing an immediate visual indication of the speed capabilities in response to an arbitrary digital input.
In this measurement the coupling strength of the monitor taper has been increased compared to the one used in \Fref{fig:profile_vs_modamp} to show the effect of the Q-factor on the switching speed.
The resonator is driven with an anti-symmetric random bit pattern clocked at the indicated bitrate. The output signal, measured at the monitor taper, is sampled synchronously and shifted such that all clock edges are overlapping at $t=0$.
Again, anti-symmetric switching means that when one input goes \low\ the other goes \high\ such that the total input and circulating powers are constant.
The switching mechanism works at a comparable speed even when modulating just one direction but the change in total circulating power and the subsequent change in the resonator temperature produce more complex dynamics.

\Fref{fig:eye}(a) shows how the system can follow the input modulation easily at a bitrate of \SI{75}{kbps}.
Similarly, at \SI{1.8}{Mbps} (b), the eye aperture is still well defined, and there is still an identifiable eye at \SI{3.8}{Mbps} (c). At bitrates of \SI{4.9}{Mbps} (d) and higher, the resonator does not complete the process of leaking out the field circulating in one direction and building up power in the opposite direction before the next switching pulse arrives. In this condition, the eye is closed and the switching reaches a steady state only if two adjacent bits are the same.
Note that, approaching the bitrate limit, the switching profile follows different paths depending on how long the resonator rested at the pre-switching state.
This is partly due to residual thermal effects caused by the transition itself and some tolerance in the symmetry between the input powers, and partly as a result of the ring-down that follows the transition (see \Fref{fig:profile_vs_modamp}).

\section{Conclusions}
The recent experimental demonstration of symmetry breaking in whispering gallery mode microresonators opens new operating regimes for the realization of bi-directional all-optical memories and switches.
The information is stored in the direction of circulation of light in the resonator and maintained as long as the counter-propagating inputs are maintained. This information is easily accessed by monitoring the transmission in counter-propagating directions and can be written by applying optical pulses with different input powers.
This principle can also be used to realize optically controlled routing of signals or simple logic devices.
The system is noise-resilient due to the thermal locking and the hysteresis that separates the two counter-propagating states. This technology may be used to implement all-optical passive signal processing on a chip without the need for conversion to and from electronic signals.

\section*{Funding}
H2020 European Research Council (ERC) (756966, CounterLight); H2020 Marie Skłodowska-Curie Actions (MSCA) (CoLiDR, 748519; GA-2015-713694); Engineering and Physical Sciences Research Council (EPSRC) (CDT for Applied Photonics and Quantum Systems Engineering Skills Hub); National Physical Laboratory (NPL) (Strategic Research).

\vfill
\section*{Acknowledgements}
L. D. B. acknowledge funding from EPSRC through the Centre for Doctoral Training in Applied Photonics.
N. M. acknowledge funding from EPSRC through the Quantum Systems Engineering Skills Hub.
We would like to thank Franc\c{o}is Copie, George N. Ghalanos, Jonathan M. Silver, Andreas \O . Svela, Michael T. M. Woodley, and Shuangyou Zhang for their helpful discussions about this work, and for reviewing this manuscript.

\bibliographystyle{apsrev4-2}
\bibliography{Leo}
\end{document}


\title{\bf Supplemental material:\\
optical memories and switching dynamics of counterpropagating light states\\
in microresonators}
\author{Leonardo~Del~Bino$^{1,2,*}$, Niall~Moroney$^{1,3,4}$, and Pascal~Del'Haye$^{1,4}$}

\affiliation{
$^{1}$National Physical Laboratory, Hampton Road, Teddington, TW11 0LW, United Kingdom\\
$^{2}$Heriot-Watt University, Edinburgh, EH14 4AS, Scotland\\
$^{3}$Imperial College London, London, SW7 2AZ, United Kingdom\\
$^{4}$Max Planck Institute for the Science of Light, Erlangen, 91058, Germany\\
$^{*}$Corresponding author: leonardo.del.bino@npl.co.uk
}
\maketitle

\section{Resonator fabrication}
The microresonator is fabricated from a 3-mm-diameter silica rod by spinning it on a spindle and using a \SI{100}{W} CO$_2$ laser to ablate the surface down to \SI{2.7}{mm} diameter and create the resonator profile. Subsequently, the surface is annealed at lower power to achieve a high Q-factor\cite{DelHaye2013}.


\section{EOM overdrive}
Being able to change the input power much faster than the response of the cavity allows us to consider the input modulation as an instantaneous change in power instead of considering the output as a convolution between the drive and the response. This simplifies the mathematical approach and the simulations, and makes the experimental results easier to understand. 
In the experiments we use an EOM to modulate the power faster than the cavity lifetime ($\gtrsim$ \SI{165}{ns}). However, EOMs respond in a particular way: about one third of the modulation follows almost instantaneously the driving voltage (\SI{10}{GHz} bandwidth), but the other two thirds are governed by charge accumulation and polarization of the electro-optic crystal that takes place at a few microseconds timescale. Since we need flat \high\ and \low\ power levels for this experiment we need to correct for this effect.
To do so, we measure the transmission of the EOMs with a Heaviside function voltage input and fit the response with an instantaneous component and three exponentially decaying terms with free time constants and weights. This fit is then used to calculate the input voltage profile that would produce the desired transmission response. This results in a rise time of \SI{8}{ns} even for full-range modulation, and a high level which is defined within \SI{2}{\%} power fluctuations after the switching.

\section{Data for simulations}
The traces in \hl{Figure~5} in the main article are generated using the parameters in \Fref{tab:data}. The Q-factor, diameter and $A_\mathrm{eff}$ for the ``SiO$_2$ rod'' are the ones measured and calculated for the resonator used in this work. For chip-based SiO$_2$ toroid resonators we use results from resonators fabricated in our group \cite{Zhang2019}. The parameters for the other materials are instead collected from other recent works (see \Fref{tab:data}).

\begin{table}[t]
\renewcommand{\arraystretch}{1.2}%
\vspace{0.5\baselineskip}
\begin{tabular}{l r r l l l}
\hline
Material    & diam.           & $A_\mathrm{eff}$       & $n$ \cite{Polyanskiy}   & $n_2$ [\si{cm^2/W}] & $Q_0$       \\
\hline
SiO$_2$ rod & \SI{2}{mm}         & \SI{50}{\micro\meter ^2}    & 1.444 & \num{2.7e-16} \cite{Kato1995}   & \num{4e8}   \\ 
SiO$_2$ toroid & \SI{100}{\micro m} & \SI{4}{\micro m^2}     & 1.444 & \num{2.7e-16}    & \num{7e7}   \\ 
Si$_3$N$_4$ \cite{Xuan2016}& \SI{600}{\micro m}  & \SI{1}{\micro m^2}     & 2.463 & \num{2.4e-15}    & \num{1.7e7}   \\ 
CaF$_2$ \cite{Savchenkov2013}    & \SI{6}{mm}  & \SI{20}{\micro m^2}     & 1.426 & \num{1.9e-16}    & \num{3e9}    \\ 
MgF$_2$  \cite{Savchenkov2013}    & \SI{2}{mm} & \SI{20}{\micro m^2}     & 1.37  & \num{9e-17}      & \num{1e9}    \\ 
Si* \cite{Yu2016}         & \SI{100}{\micro m}  & \SI{1}{\micro m^2}     & 3.43  & \num{1.7e-14} \cite{Lin2007}   & \num{7e5} \\
\bottomrule
\end{tabular}
\caption{Parameters used for the simulation in Figure~6 in the main article. * For silicon a wavelength of \SI{3.1}{\micro m} is used instead of \SI{1.55}{\micro m}.
$A_\mathrm{eff}$ is the effective mode area, $n$ is the refractive index, $n_2$ is the nonlinear refractive index, and $Q_0$ is the intrinsic Q-factor.}
\label{tab:data}
\end{table}

\section{Other parameters affecting the switching}

\begin{figure}[bth]
\centering
  \includegraphics[width=\columnwidth]{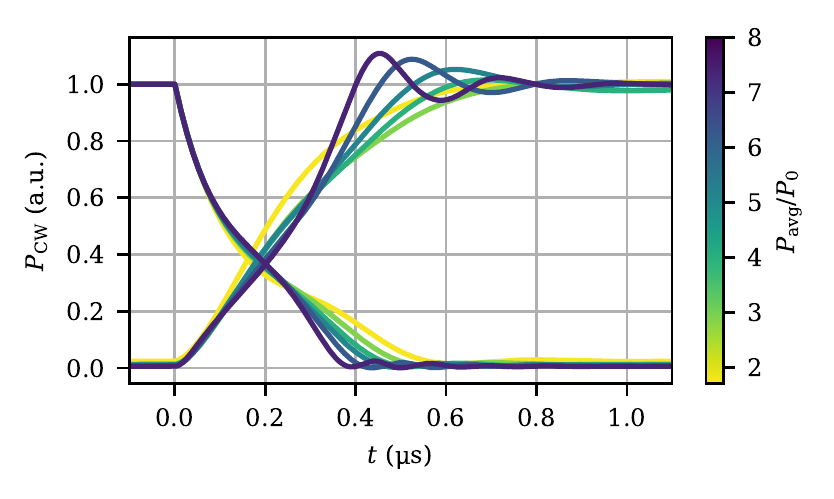}
\caption{Simulation of the switching profile for different input powers. The average input power range from $P_\mathrm{in} = 1.7P_0$ to $P_\mathrm{in} = 8P_0$.}
    \label{fig:switch-vs-pow}
\end{figure}

The modulation amplitude is the most important parameter affecting the switching speed, but the laser detuning and the total input power also have a small effect on the switching profile. \Fref{fig:switch-vs-pow} shows how the power affects the switching profile. The rise time varies by about \SI{30}{\%} over the range of input powers considered. In addition, an overshoot and ringdown arises at higher powers with amplitude and frequency of the overshoot increasing as the power increases. 

\section{Input signals}

\begin{figure}[bt]
\centering
  \includegraphics[width=\columnwidth]{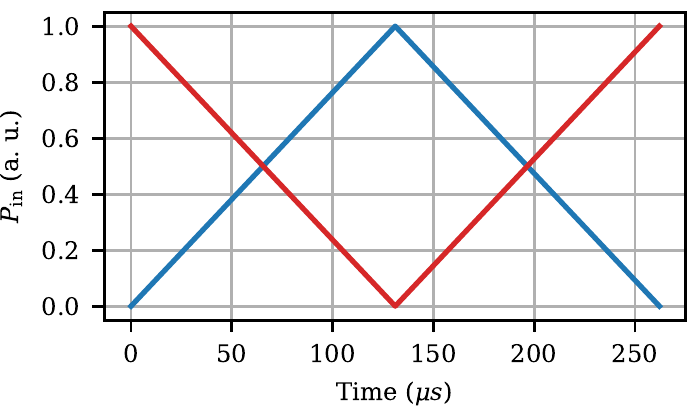}
\caption{Input power profile in the two directions (blue and red) used in Figure~3 in the main article.}
    \label{fig:Pin-hyst}
\end{figure}

\Fref{fig:Pin-hyst} shows the input powers used to measure the hysteresis profiles displayed in Figure~3 in the main article. Note that the total power sent to the resonator is constant and the ramp taking place on a timescale much longer than the switching speed.
\hl{The green curve in Figure~3(g) in the main article is calculated by running the time step simulation at several input powers and taking the maximum hysteresis amplitude for each of them. The simulation parameters such as Q-factor, resonator diameter, wavelength are the ones measured for the resonator used in the experiment. The value of $P_0$ is calculated from the minimum input power to observe symmetry braking via Equation~(7) in the main article.}

\begin{figure}[tb]
\centering
  \includegraphics[width=\columnwidth]{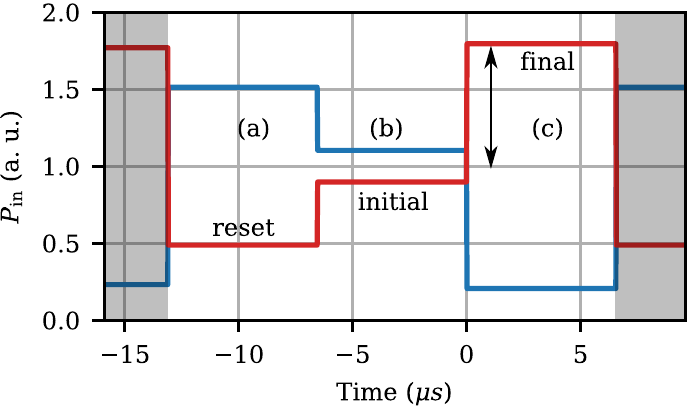}
\caption{A cycle of the input power profile in the two directions (blue and red) used in Figure~4 in the main article. The origin of the time scale is set accordingly. The three phases of the cycle are marked and the variable part is highlighted by an arrow.}
    \label{fig:Pin-switch}
\end{figure}

The input power used to measure and simulate the switching profile in Figure~4 in the main article is shown in \Fref{fig:Pin-switch}.
Each cycle starts with a reset phase (a) when the power is imbalanced enough to overcome the hysteresis and bring the red direction to the \low\ state.
The powers are then returned to near equality (10 \% imbalanced) and this constitutes the initial state (b).
The powers are then imbalanced to the final state (c) and the temporal profile of the coupled power in the red direction is measured.
The amount of imbalance in the final phase (indicated by the arrow in \Fref{fig:Pin-switch}) is varied from 0 to 100~\% of the average power over 200 cycles.
Some of the cycles are plotted in Figure~4 in the main article with the relative imbalance shown by the color of the line.

\begin{figure}[bth]
\centering
  \includegraphics[width=\columnwidth]{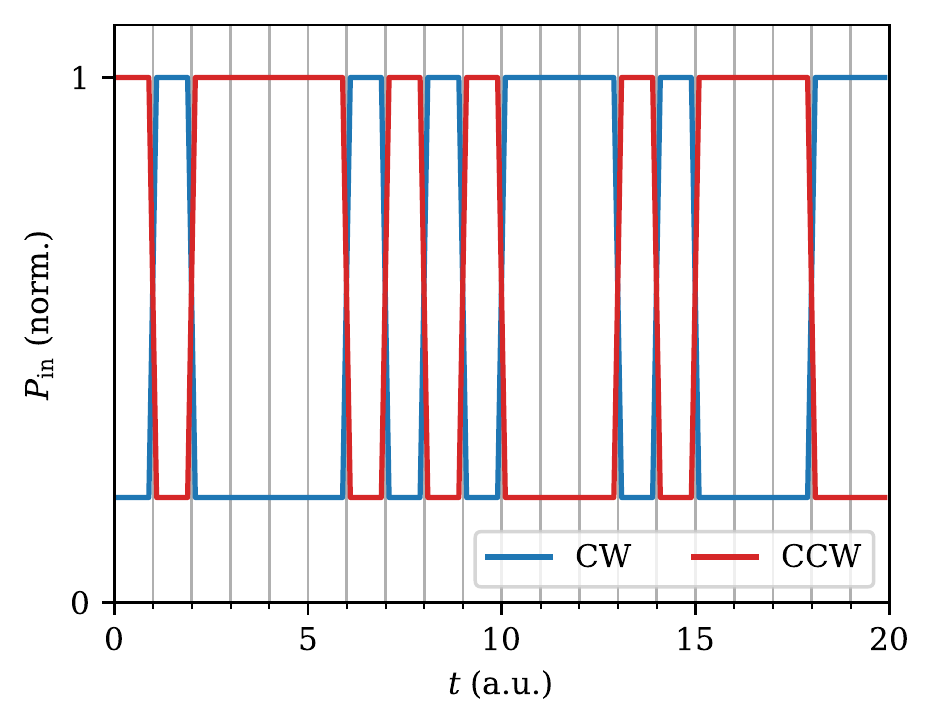}
\caption{An example of random bit sequence input signal used to measure Figure~6 in the main paper.}
    \label{fig:Pin-eye}
\end{figure}

\hl{To create the eye diagrams shown in Figure~6 in the main article we use a random bit input as the one shown in} \Fref{fig:Pin-eye}.
\hl{The clock transitions of the input signal are represented by the vertical grid.
For each clock transition the signal can randomly switch configuration or stay constant.
The bitrate is given by the frequency of the clock.
The input power in the two directions is modulated anti-symmetrically to conserve the total power to the resonator. We chose a total input power high enough to create hysteresis between the two states and a power imbalance between the two directions high enough to achieve the fastest switching speed possible.
The output signal is then plotted multiple times in Figure~6 in the main article. Each time, the time-scale is offset so that each clock transition is overlapped at $t=0$. This allows to compare all the possible transitions at the output on a zoomed in time-scale.}

\section{Coupling and Q-factor}
To obtain the curves in \hl{Figure~5} in the main article, we compare the threshold power $P_\mathrm{th}$ defined in \Fref{eq:Pth}, and the bitrate $R$ defined in the main article.
The coupled Q-factor is part of both equations. It is defined from the intrinsic linewidth $\gamma_0$ and the coupling strength $\kappa$ as follows:
\begin{equation}
Q_0 =  \frac{\omega}{2 \, \gamma_0}; \qquad Q =  \frac{\omega}{2\gamma}.
\end{equation}
with $\nu = \omega /2 \pi$ being the optical frequency. The coupling efficiency, i.e.~the maximum fraction of light that can get transferred from the tapered fiber to the resonator, is given by
\begin{equation}
\eta=\frac{4 \, \kappa \, \gamma_0}{\gamma^2},
\end{equation}
where the coupled linewidth is $\gamma = \kappa+\gamma_0$.
The maximum coupling efficiency $\eta = 1$ is obtained for $\kappa=\gamma_0$.

For each type of resonator a compromise between speed and power consumption is chosen by tuning the coupling of the resonator to the input/output waveguide. In \hl{Figure~5} in the main article, the parameter $\kappa$ varies from $0.03\gamma_0$ to $30\gamma_0$ and the corresponding threshold power, and bitrate are plotted using the following equations.
\begin{equation}
P_\mathrm{th} = \frac{1.54}{\eta}\frac{\pi^2 \, n_0^2 \, d \, A}{n_2 \, \lambda \, Q(\kappa) \, Q_0},
\label{eq:Pth}
\end{equation}
\begin{equation}
R = \frac{1}{2.197}\frac{\omega}{Q(\kappa)}
\end{equation}
\bibliographystyle{apsrev4-2}
\bibliography{Leo}